\documentclass[12pt, aps, prc, preprintnumbers, tightenlines,superscriptaddress, nofootinbib, float]{revtex4-1}

\usepackage[utf8]{inputenc}
\usepackage{graphicx} 
\usepackage[dvipsnames]{xcolor}
\usepackage{amsmath,amssymb,float}
\usepackage{caption}
\usepackage{subcaption}
\usepackage{ragged2e}  
\DeclareCaptionJustification{justified}{\justifying}
\definecolor{lcolor}{rgb}{0.5,0,0}
\definecolor{citcolor}{rgb}{0,0.3,0.0}
\usepackage{url}
\usepackage[breaklinks,colorlinks,citecolor=citcolor,urlcolor=blue,linkcolor=lcolor]{hyperref}
\usepackage[capitalise]{cleveref}
\usepackage{wrapfig}
\usepackage{comment}
\usepackage{blindtext}
\newcommand\pubdate{\today}

\newcommand{\lqcd}{\Lambda_{\mathrm{QCD}}}

\newcommand{\xpom}{x_{I\hspace{-0.3em}P}}
\newcommand{\gev}{\mathrm{GeV}}
\newcommand{\xbj}{{x}}
\newcommand{\ygap}{Y_{\rm gap}}

\newcommand{\rt}{{\mathbf{r}}}
\newcommand{\ra}{{\mathbf r}_1}
\newcommand{\rb}{{\mathbf r}_2}
\newcommand{\bt}{{\mathbf{b}}}
\newcommand{\ba}{{\mathbf{b}}_1}
\newcommand{\bb}{{\mathbf{b}}_2}
\newcommand{\dd}{{\mathrm{d}}}

\def\Title#1{\begin{center} {\Large #1 } \end{center}}
\def\Author#1{\begin{center}{ \sc #1} \end{center}}
\def\Address#1{\begin{center}{ \it #1} \end{center}}

\newcommand\pubblock{\rightline{\begin{tabular}{l}  \\ % Author's note number [if you need to add one] goes here
         \pubdate  \end{tabular}}}
\newenvironment{Abstract}{\begin{quotation}  }{\end{quotation}}
\newenvironment{Presented}{\begin{quotation} \begin{center} 
             PRESENTED AT\end{center}\bigskip 
      \begin{center}\begin{large}}{\end{large}\end{center} \end{quotation}}
\begin{document}
\begin{titlepage}
 \pubblock
\vfill
\Title{Coherently diffractive dissociation in electron-hadron collisions: from HERA to the future EIC}
\vfill
\Author{Tuomas Lappi$^{1,2}$, Anh Dung Le$^{1,2}$, and Heikki Mäntysaari$^{1,2}$}
\Address{$^{1}$Department of Physics, University of Jyväskylä, P.O. Box 35, 40014 University of Jyväaskylä, Finland}
\Address{$^{2}$Helsinki Institute of Physics, P.O. Box 64, 00014 University of Helsinki, Finland}
\vfill
\begin{Abstract}
We present numerical results on diffractive dissociation with large invariant mass diffractive
final states in the scattering of an electron off a hadron. The diffractive large-mass resummation is performed using the nonlinear Kovchegov-Levin equation, taking into account  running coupling corrections. For the scattering off the proton, a (modified) McLerran-Venugopalan amplitude is used as the initial condition for the nonlinear evolution, with free parameters being constrained by the HERA inclusive data. The results show a reasonable description of the HERA diffractive structure function data at moderately large diffractive mass when the impact parameter profile is constrained by the low-mass diffractive cross section data. The calculation is extended to nuclear scattering, where the initial condition is generalized from the proton case employing the optical Glauber model. The nonlinear large-mass resummation predicts a strong nuclear modification in diffractive scattering off a nuclear target in kinematics accessible at the future Electron-Ion collider. 
\end{Abstract}
\vfill
\begin{Presented}
DIS2023: XXX International Workshop on Deep-Inelastic Scattering and
Related Subjects, \\
Michigan State University, USA, 27-31 March 2023 \\
     \includegraphics[width=9cm]{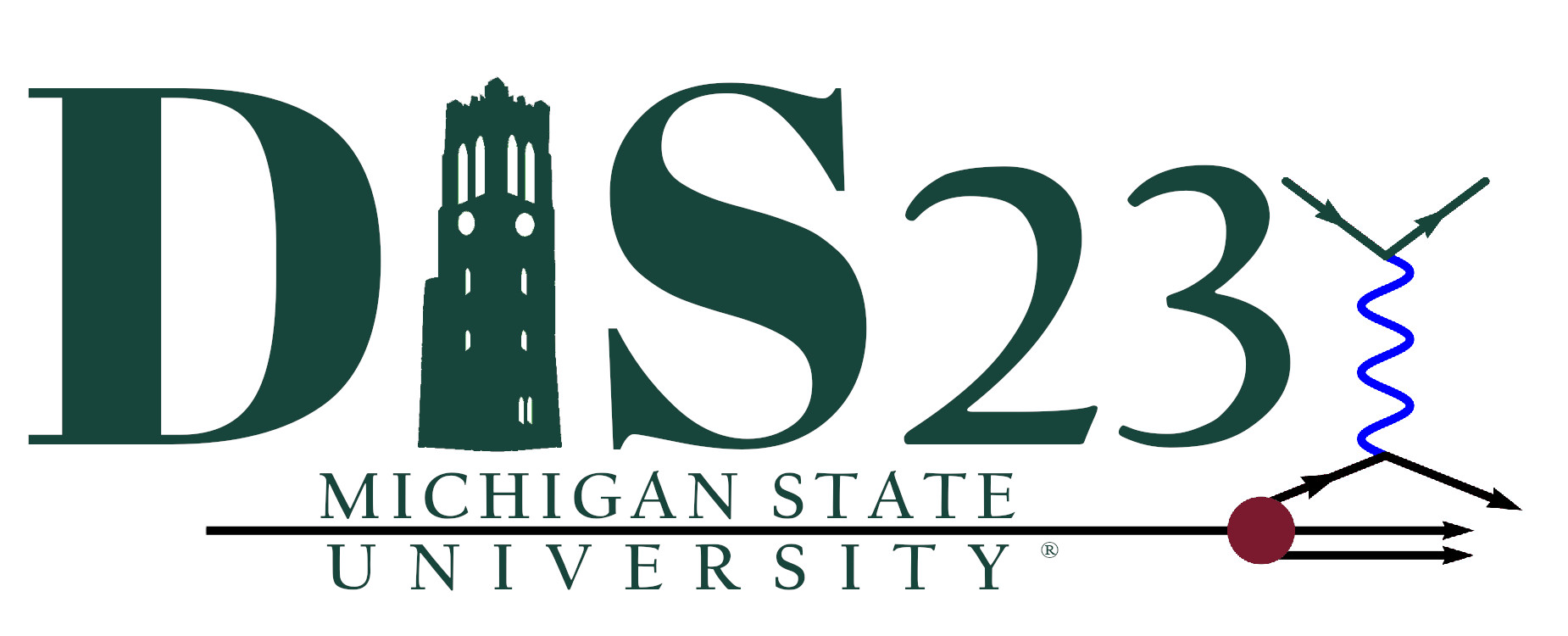}
\end{Presented}
\vfill
\end{titlepage}

\section{Introduction}

The study of diffractive processes in deeply inelastic electron-hadron scattering is a longstanding topic in the high-energy particle physics. It was impressively boosted by the striking experimental signature in electron-proton collisions at  HERA~\cite{H1:1995cha,ZEUS:1995sar}, and by the observation that  diffraction is an excellent probe for nonlinear gluon saturation thanks to the color-singlet nature of the $t$-channel exchange (pomeron) which can be described by two gluon exchange at lowest order.

In deep-inelastic scattering, both diffractive and inclusive processes can be conveniently described using the dipole picture (see e.g. Ref.~\cite{Kovchegov:2012mbw} for a review) %~\cite{Kopeliovich:1981pz,Mueller:1989st,Nikolaev:1990ja} 
in which the energetic virtual photon mediating the interaction is replaced by its $|q\Bar{q}\rangle$ Fock state at lowest order. In this picture, the total and diffractive virtual photon-hadron cross sections can be factorized schematically as
\begin{equation}
\label{eq:dipole_factorization}
    \sigma_{\rm tot}^{\gamma^*h} = |\Psi^{\gamma^*\to q\Bar{q}}|^2_{L+T} \otimes \sigma^{q\Bar{q}h}_{\rm tot} \quad {\rm and}\quad \frac{\dd\sigma_{\rm D}^{\gamma^*h}}{\dd\ln(1/\beta)} = |\Psi^{\gamma^*\to q\Bar{q}}|^2_{L+T} \otimes \frac{\dd\sigma^{q\Bar{q}h}_{\rm D}}{\dd\ln(1/\beta)}, 
\end{equation}
respectively, after integrating out the impact parameter dependence. The convolutions indicate the integration over all possible $q\Bar{q}$-dipole configurations whose probability is given by the squared wave functions $|\Psi^{\gamma^*\to q\Bar{q}}|^2_{L,T}$ in the longitudinal (L) and transverse (T) components (see e.g. Ref.~\cite{Kovchegov:2012mbw} for details). Here diffraction is limited to the coherent case, in which the target remains intact after scattering.
The  inclusive process can be characterized using the photon virtuality $Q^2$ and the Bjorken variable $\xbj$, while the diffractive process can be conveniently described by the triplet $(Q^2,\beta,\xpom)$. Here $\beta$ is related to the invariant mass $M_X$ of the diffractively produced system, $\beta = \ln[Q^2/(Q^2+M_X^2)]$, and $\xpom = \xbj/\beta$ encodes the size of the rapidity gap, $\ygap = \ln(1/\xpom)$, which is a consequence of the exchange of vacuum quantum numbers and an important signature of  diffractive events. 
%In practice, the diffractive structure functions $F_{2,L,T}^{D(3)}$ and reduced cross section $\sigma_{\rm red}^{D(3)}$ are usually used, which are related to the diffractive cross-section as
%
%\begin{subequations}
%\begin{equation}
%    F_{L,T}^{D(3)} = \frac{Q^2}{4\pi^2\alpha_{em}} \left[\frac{\dd\sigma_{\rm D}^{\gamma^*h}}{\dd\ln(1/\beta)}\right]_{L,T}, \quad F_2^{D(3)} = F_L^{D(3)} + F_T^{D(3)},
%\end{equation}
%
%\begin{equation}
%    \sigma_{\rm red}^{D(3)} = F_2^{D(3)} - \frac{y^2}{1+(1-y)^2} F_L^{D(3)},
%\end{equation}
%\end{subequations}
%
%with $y=Q^2/(xs)$ the inelasticity of the scattering at center-of-mass energy $\sqrt{s}$.

The gluon saturation can be incorporated naturally in the dipole picture using the Color Glass Condensate (CGC) effective theory (for a review, see e.g. Ref.~\cite{Gelis:2010nm}) in which the dipole-hadron cross sections can be obtained from the solution to the nonlinear evolution equations. At large to medium $\beta$, when it is sufficient to consider only the $q\Bar{q}$ and $q\Bar{q}g$ components of the photon wave function~\cite{Munier:2003zb,Kugeratski:2005ck,Marquet:2007nf,Kowalski:2008sa, Kowalski:2008sa,Cazaroto:2008iy,Bendova:2020hkp, Beuf:2022kyp}, the diffractive cross-sections can be expressed in terms of the solution to the so-called Balitsky-Kochegov (BK) equation~\cite{Balitsky:1995ub,Kovchegov:1999yj,Kovchegov:2006vj,Balitsky:2006wa}. 
Going down to small $\beta$, one should resum the soft-gluon contributions by the so-called Kovchegov-Levin (KL) evolution equation or by one of its generalizations~\cite{Kovchegov:1999ji,Kovner:2001vi,Hentschinski:2005er,Kovner:2006ge,Levin:2001yv,Levin:2001pr,Levin:2002fj,Hatta:2006hs,Kovchegov:2011aa,Lublinsky:2014bma,Contreras:2018adl,Le:2021afn}. Extending calculations to the small-$\beta$ region, together with achieving higher-order accuracy, is crucial in both theoretical and experimental aspects. This is especially true for the investigation of diffractive dissociation events at the future DIS facilities, such as EIC~\cite{AbdulKhalek:2021gbh} and LHeC/FCC-he~\cite{LHeC:2020van}.                               

In this study, we focus on diffractive DIS with moderately high mass, $\beta\lesssim 0.1$, by the means of the Kovchegov-Levin equation. The evolution requires some non-perturbative input at moderately small $\xbj$, which is constrained from the HERA inclusive structure function data. The scatterings off the proton and off the nucleus are considered separately, and we present comparisons to the HERA data and predictions for the nuclear diffractive cross section to be measured at the EIC. %which are dedicated for a description of the HERA diffractive data and for a prediction of diffractive dissociation at EIC in the desired regime. 
This contribution is based on the work presented in Ref.~\cite{Lappi:2023frf}. 

\section{Kovchegov-Levin formalism for diffractive dissociation}
\label{sec:formalism}

The basic ingredient of dipole factorization~(\ref{eq:dipole_factorization}) is the dipole-target cross section. 
Let us consider the scattering of a dipole of transverse size $\rt$ off a target hadron at impact parameter $\bt$ and at the total relative rapidity $Y=\ln(1/\xbj)$. At a large number of colors ($N_c$) and in the eikonal limit, the forward elastic scattering amplitude $N(\rt,Y;\bt)$ evolves in rapidity according to the BK equation 
\begin{multline}
\label{eq:Bk_equation}
    \partial_Y N(\rt,Y,\bt) = \int \dd^2\ra\ \mathbb{K} (\rt,\ra,\rb) \left[N(\ra,Y;\ba) + N(\rb,Y;\bb) \right.\\ \left. - N(\rt,Y;\bt) - N(\ra,Y;\ba)N(\rb,Y;\bb)\right],
\end{multline}
starting from a certain initial input at some $\xbj = \xbj_{\rm init}$ (to be specified in the next section), where $\rb = \rt - \ra$, $\ba = \bt - (\rb/2)$ and $\bb = \bt + (\ra/2)$. The integral kernel $\mathbb{K} $ is some function of the transverse sizes whose form depends on the considered scenario. In this analysis, the running-coupling effect is involved by adopting the Balitsky's prescription from Ref.~\cite{Balitsky:2006wa}.
%
%\begin{equation}
%\label{eq:Bal_presc}
%    \mathbb{K}^{\rm rc}_{\rm Bal} (\rt,\ra,\rb) = \frac{N_c\alpha_s(\rt^2)}{2\pi^2} \left[ \frac{\rt^2}{\ra^2\rb^2} + 
%    \frac{1}{\ra^2}\left(\frac{\alpha_s(\ra^2)}{\alpha_s(\rb^2)}-1\right) \right.\\ 
%    \left. + \frac{1}{\rb^2}\left(\frac{\alpha_s(\rb^2)}{\alpha_s(\ra^2)}-1\right)\right].
%\end{equation}
%
The strong coupling in coordinate space runs with the squared dipole size $\rt^2$,
\begin{equation}    \label{eq:running_coupling}
    \alpha_s(\rt^2) = \frac{12\pi}{(33-2N_f)\ln \frac{4C^2}{\rt^2\lqcd^2}}~ \Theta (r\leq r_{c}) + 0.7~\Theta (r> r_{c}), 
\end{equation}
where $r \equiv |\rt|$ and $r_c$ solves $\alpha_s(r_c^2) = 0.7$. We include the second term in~\cref{eq:running_coupling} to avoid the Landau pole. The constant $C^2$ accounts for the uncertainty when transforming from momentum space to coordinate space.
%, and is obtained from a fit to proton inclusive structure function data e.g. in Refs.~\cite{Lappi:2013zma,Albacete:2010sy}, together with the non-perturbative input of the BK evolution. Here we use the fits in Ref.~\cite{Lappi:2013zma}, and consequently adopt the same setup and work with only light quarks ($N_f=3, m_f=140\,\mathrm{MeV}$). The detailed setups are discussed in~\cref{sec:results} for the scattering off both proton and nucleus. 
The total dipole-target cross section can be deduced from the forward amplitude $N$ using the optical theorem:
\begin{equation}
    \sigma_{\rm tot}^{q\Bar{q}h} (\rt,Y) = 2\int \dd^2\bt N(\rt,Y,\bt).
\end{equation}

Let us now turn to the discussion of diffraction. At small $\beta$, the parameter $\alpha_s \ln (1/\beta)$ associated with the contribution of one-gluon emission becomes large, which requires a resummation. At large $N_c$, this resummation is doable using the KL equation. Denoting by $N_{\rm D}(\rt,Y,Y_0;\bt)$ the diffractive dipole-target cross section with a \emph{minimal} rapidity gap $Y_0$, the KL evolution is basically the BK evolution (\ref{eq:Bk_equation}) for the auxiliary function $N_I \equiv 2N(\rt,Y;\bt) - N_{\rm D}(\rt,Y,Y_0;\bt)$. The initial condition is given at $Y=Y_0$ at which the scattering is quasi-elastic as $N_{\rm D}(\rt,Y_0,Y_0;\bt) = N^2(\rt,Y_0;\bt)$. Knowing $N_D$, the diffractive dipole-hadron cross section appearing in~\cref{eq:dipole_factorization} can be obtained straightforwardly as
\begin{equation}
    \frac{\dd\sigma^{q\Bar{q}h}_{\rm D}}{\dd\ln(1/\beta)} = \int d^2\bt \left .\left(-\frac{\dd N_{\rm D}(\rt,Y,Y_0,\bt)}{\dd Y_0}\right) \right|_{Y_{0} = \ygap}.
\end{equation}

It is constructive to mention also low-mass diffraction in brief (for the complete formulation, see e.g. Ref.~\cite{Kowalski:2008sa}). At medium values of beta, $0.1\sim\beta<0.5$, 
the diffractive scattering is dominated by the $q\Bar{q}g$ contribution.
%, and in the limit of large $Q^2$, the diffractive structure function is schematically proportional to the dipole cross section in the adjoint representation, $ F^{D(3)}_{q\Bar{q}g} \sim 2N(\rt,Y_{gap};\bt) - N^2(\rt,Y_{gap};\bt)$.
The $q\Bar{q}$ component becomes more important when going up in $\beta$, and it dominates at $\beta>0.5$. Its contribution is proportional to the dipole elastic cross section, $ F^{D(3)}_{q\Bar{q}} \sim N^2(\rt,Y_{gap};\bt)$. We shall use the latter contribution as a reference to fit the proton shape, see~\cref{sec:results}.

\section{Results}
\label{sec:results}
\subsection{Electron-proton scattering}
\label{subsec:ep}

In the scattering off the proton, we assume a simple impact-parameter factorization of the dipole profiles, which reads
\begin{subequations}
    \begin{equation}
        \label{eq:factorize_N}
        \sigma^{q\Bar{q}p}_{\rm tot} = 2\int \dd^2\bt N(\rt, Y; \bt) = \mathcal{N}(r,Y)\times 2\int d^2\bt T_p(\bt) = \sigma_0 \mathcal{N}(r,Y), 
    \end{equation}
    \begin{equation}
        \label{eq:factorize_ND}
        \sigma^{q\Bar{q}p}_{\rm D} = \int \dd^2\bt N_{\rm D}(\rt, Y, Y_0; \bt) = \mathcal{N}_{\rm D}(r,Y,Y_0)\times \int d^2\bt T_p^2(\bt) = \sigma^{\rm D}_0 \mathcal{N}_{\rm D}(r,Y,Y_0).
    \end{equation}
\end{subequations}
\begin{wrapfigure}{l}{0.62\textwidth}
\setlength\intextsep{0pt}
  \centering
    \includegraphics[width=0.58\textwidth]{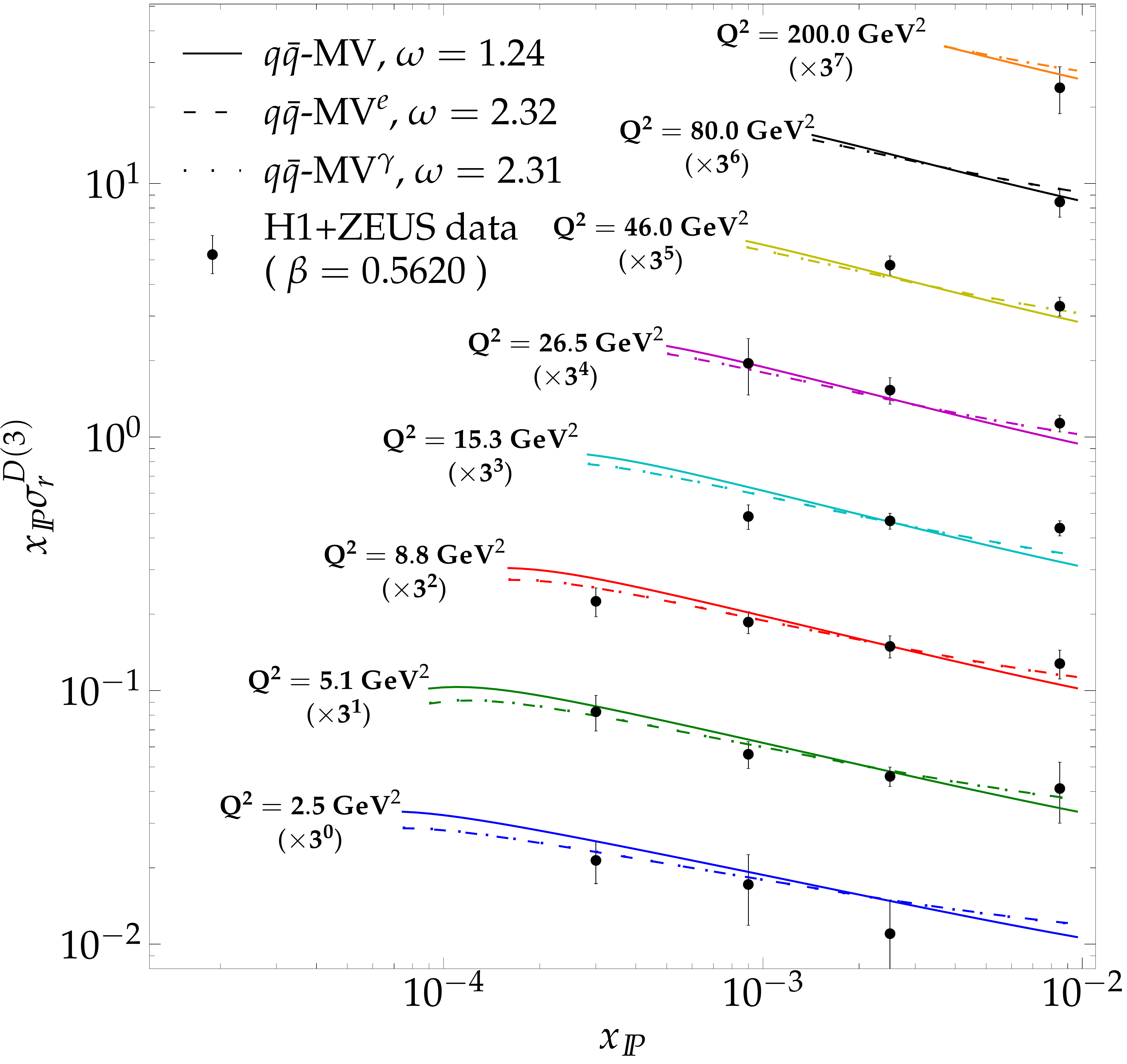}
  \caption{The reduced diffractive cross sections taking into account only the $q\Bar{q}$ contribution fitted to the HERA combined data~\cite{H1:2012xlc} at $\beta>0.5$ (here, only $\beta=0.562$ are shown) and at different $Q^2$ bins. The optimal values for $\omega$ correspoding to different parametrizations are shown in the legend.}
  \label{fig:gbw_fit}
\end{wrapfigure}
Here the $\bt$-independent functions $\mathcal{N}(r,Y)$ and $\mathcal{N}_{\rm D}(r,Y,Y_0)$ obey the $\bt$-independent BK and KL equations, respectively. \cref{eq:factorize_ND} follows from the initial condition for the KL evolution. 
The proton density profile $T_p(\bt)$ is chosen to be the regularized gamma function, 
\begin{equation}
    T_p(\bt) = \frac{\Gamma\left(\frac{1}{\omega},\frac{\pi b^2}{\omega(\sigma_0/2)}\right)}{\Gamma\left(\frac{1}{\omega}\right)},
    \label{eq:b-profile-proton}
\end{equation}
where $\omega \geq 0$ controls the steepness of the profile. Note that when $\omega=1$, it becomes the usual gaussian density. The normalization of the diffractive cross section $\sigma^{\rm D}_0$ is particularly sensitive to this parameter. 

%\begin{table*}[h!] 
%        \centering
%        \begin{tabular}{l | c c c c c | c} 
%        \hline
%        Parametrization & $Q_{s0}^2 (\rm GeV^2)$ & $\gamma$ & $e_c$ & $\sigma_0/2\ ({\rm mb})$ & $C^2$ & $\ \omega_\mathrm{opt}$ \\ [0.5ex] 
%        \hline\hline
%        MV & 0.104 & 1 & 1 & 18.81 & 14.5 &\ 1.24\\ 
%        MV$^e$ & 0.060 & 1 & 18.9 & 16.36 & 7.2 &\ 2.32 \\ 
%        MV$^\gamma$ & 0.159 & 1.129 & 1 & 16.35 & 7.05 &\ 2.31 \\[1ex]
%        \hline
%        \end{tabular}
%        \caption{Parameters from different fits used in the calculation (from Refs.~\cite{Lappi:2013zma}). The determined optimal values for the parameter $\omega$ in Eq.~\eqref{eq:b-profile-proton} are also shown. }
%        \label{tab:params}
%\end{table*}

Based on Ref.~\cite{Lappi:2013zma}, the following parametrization is taken as the initial condition for the BK evolution at $x_{\rm init} = 0.01$:
\clearpage
\begin{equation}
    \label{eq:mvparam}
    \mathcal{N}_{MV}(r) = 1 - \exp\left[-\frac{(r^2Q_{s0}^2)^{\gamma}}{4}\ln\left(e\cdot e_c + \frac{1}{r\lqcd}\right)\right],
\end{equation}
%
%
%
%\clearpage
where $Q_{s0}$ controls the initial saturation scale, $\gamma$ is the anomalous dimension and $e_c$ regularize the large-$r$ behavior. Their values, together with the two above-mentioned free parameters $C^2$ and $\sigma_0$, used in this analysis are taken from the fits to the HERA inclusive structure function data~\cite{H1:2009pze} using only the light quarks reported in Ref.~\cite{Lappi:2013zma}. We shall use also the notations MV, MV$^e$ and MV$^\gamma$ for the three different fits there in this study.
%The normalization factor $\sigma_0$ for the total cross section can be obtained from a fit to the proton inclusive structure function data, and once it is fixed, the normalization $\sigma_D$ of the diffractive cross section will be sensitive on the proton density profile $T_p(\bt)$. We choose the following  

The only remaining free parameter is the steepness $\omega$. To estimate its optimal value, we choose to fit the $q\Bar{q}$ contribution only to the HERA diffractive combined data~\cite{H1:2012xlc} with $\beta>0.5$ ($24$ data points). We obtained good fits, giving $\omega_\mathrm{opt} \simeq 1.24\ (\chi^2/{\rm dof} \approx 1.87)$ for MV, $\omega_\mathrm{opt} \simeq 2.32\ (\chi^2/{\rm dof} \approx 1.08)$ for MV$^{e}$, and $\omega_{opt} \simeq 2.31\ (\chi^2/{\rm dof} \approx 1.09)$ for MV$^{\gamma}$ (see~\cref{fig:gbw_fit}). 
%The proton profiles resulted from these values are significantly steeper than the usual gaussian shape.

\begin{figure}[ht!]
    \centering
        \begin{subfigure}[b]{0.491\textwidth}
         \centering
         \includegraphics[width=\textwidth]{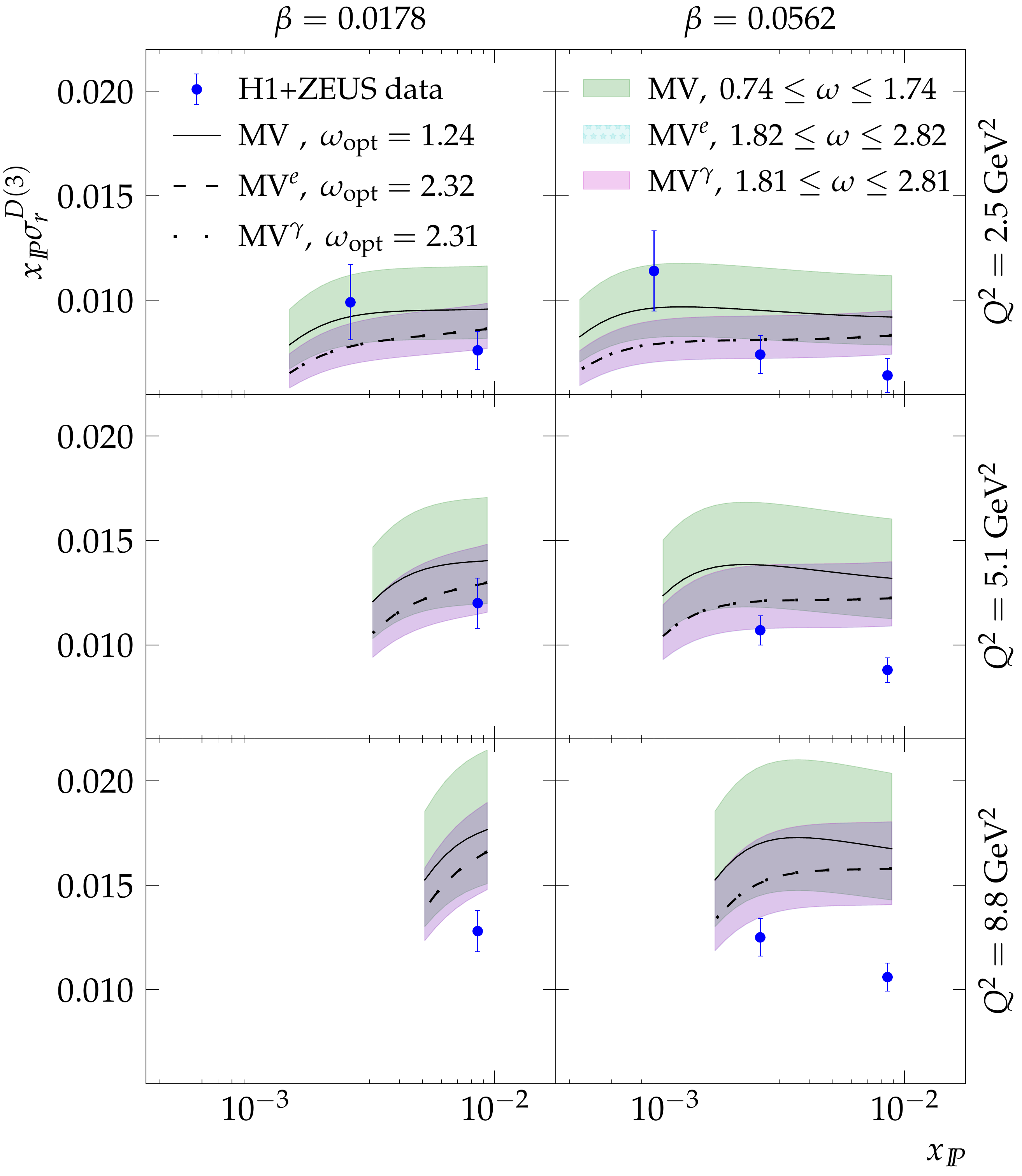}
         \caption{$\xpom$ dependence}
         \label{fig:KL_HERA_comparison_xP}
        \end{subfigure}
        \hfill
        \begin{subfigure}[b]{0.491\textwidth}
         \centering
         \includegraphics[width=\textwidth]{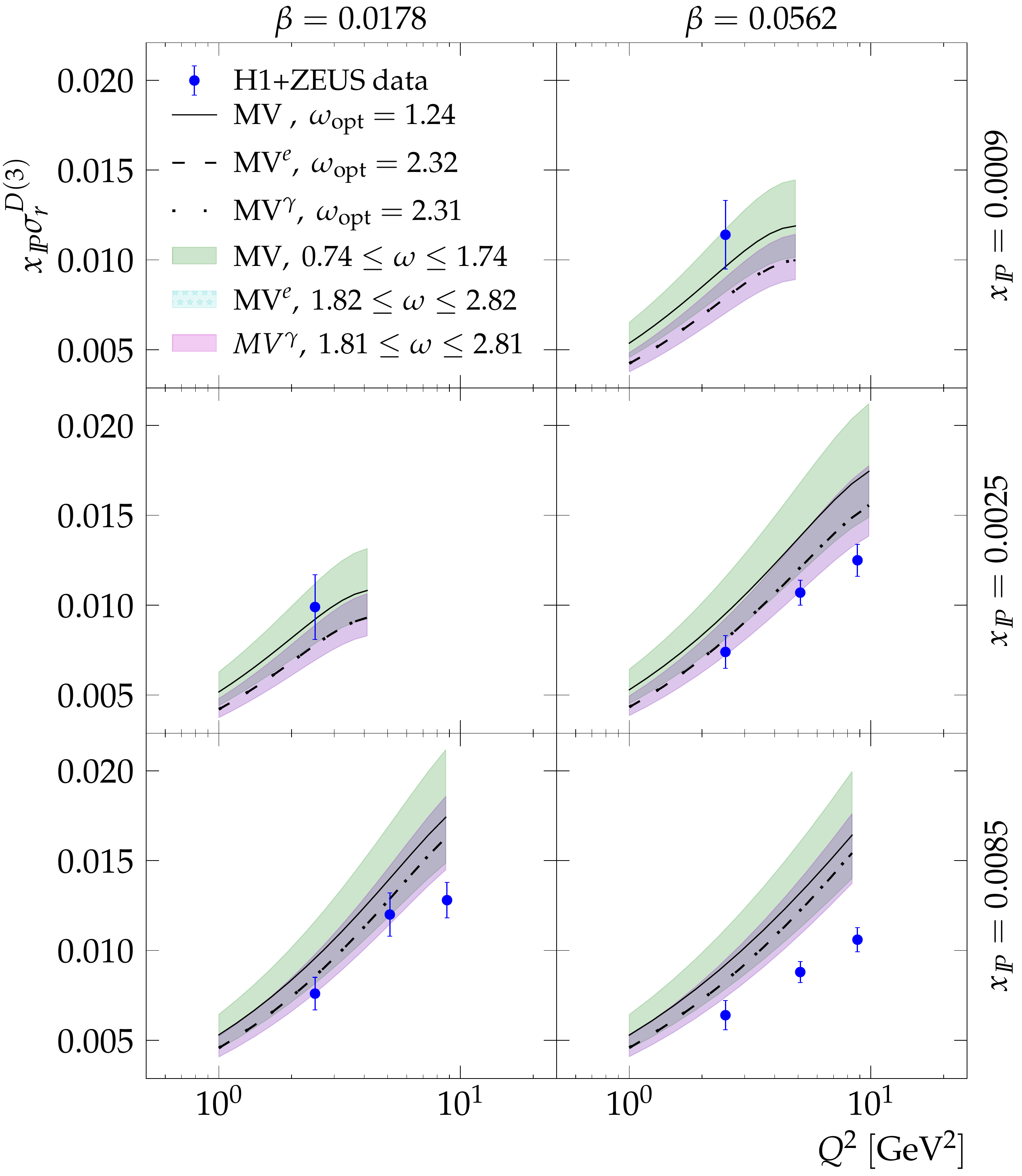}
         \caption{$Q^2$ dependence}
         \label{fig:KL_HERA_comparison_Q2}
        \end{subfigure}
        \caption{Comparison to the HERA combined data~\cite{H1:2012xlc} at moderately small $\beta<0.1$ and $Q^2<10 ~\gev^2$. The lines correspond to the numerical KL results at the optimal values of $\omega$, while the bands represent the variation of the numerics for $\omega_{\rm opt} - 0.5 \leq \omega \leq \omega_{\rm opt} + 0.5$.}
        \label{fig:hera_comparison}
\end{figure}

A comparison to the HERA combined data on the reduced diffractive cross section using the numerical solutions to the Kovchegov-Levin equation, with the above-obtained values of $\omega_{\rm opt}$, for a narrow kinematics range of $\beta<0.1$ and $Q^2<10 ~\gev^2$ is shown in~\cref{fig:hera_comparison}. Overall, the numerics shows a good agreement to the data at such a moderately small $\beta$, though the normalization is typically overestimated when going up in $Q^2$. The mild $\xpom$ dependence seen at those small virtualities is compatible with the predictions from  KL evolution. Meanwhile,  KL evolution predicts an evident rise towards higher $Q^2$, which could be seen in the HERA combined data. A more complete comparison is available in Ref.~\cite{Lappi:2023frf}.

\subsection{Electron-nucleus scattering}
\label{subsec:eA}

Let us consider now the scattering off a nuclear target $A$, inheriting some knowledge on the electron-proton scattering from the above discussions. Instead of assuming that the impact parameter profile factorizes as in \cref{eq:factorize_N,eq:factorize_ND}, we now include the impact parameter $b=|\bt|$ dependence in the dipole-nucleus amplitude and solve the evolution equations at each impact parameter independently. The initial condition at $x_{\rm init} = 0.01$ now reads:
\begin{equation}
    \label{eq:glauber_mv_ic}
    N_A(r,b) =  1 - \exp\left[-AT_A(b)\frac{\sigma_0}{2}\frac{(r^2Q_{s0}^2)^{\gamma}}{4}\ln\left(e\cdot e_c + \frac{1}{r\lqcd}\right)\right],
\end{equation}
which is obtained from \cref{eq:mvparam} using the optical Glauber model following Ref.~\cite{Lappi:2013zma}. The nuclear density function $T_A(b)$ is taken to be the Wood-Saxon parametrization, with the parameters specified in the same cited reference. The values of all other parameters are adopted from the three fits discussed in \cref{subsec:ep}.

Following also Ref.~\cite{Lappi:2013zma} we note a rapid increase of the gluon density in the low density
region, $b > b_{\rm cut} \approx 6.3 \rm~ fm$, which would lead to unphysically rapid growth
of the nuclear size. In such regime, we assume the following impulse approximation instead of using the solutions to the nonlinear evolutions for the nuclear target:
\begin{subequations}
    \label{eq:large_b_cut}
    \begin{equation}
        N_A(r,Y;b>b_{\rm cut}) = (\sigma_0/2)AT_A(b) \mathcal{N}(r,Y),    
    \end{equation}
    \begin{equation}
        N_{{\rm D},A}(r,Y,Y_0;b>b_{\rm cut}) = (\sigma_0^2/4)A^2T_A^2(b)\mathcal{N}_{\rm D}(r,Y,Y_0).   
    \end{equation}
\end{subequations}

\begin{figure}[h!]
    \centering
    \includegraphics[width=\textwidth]{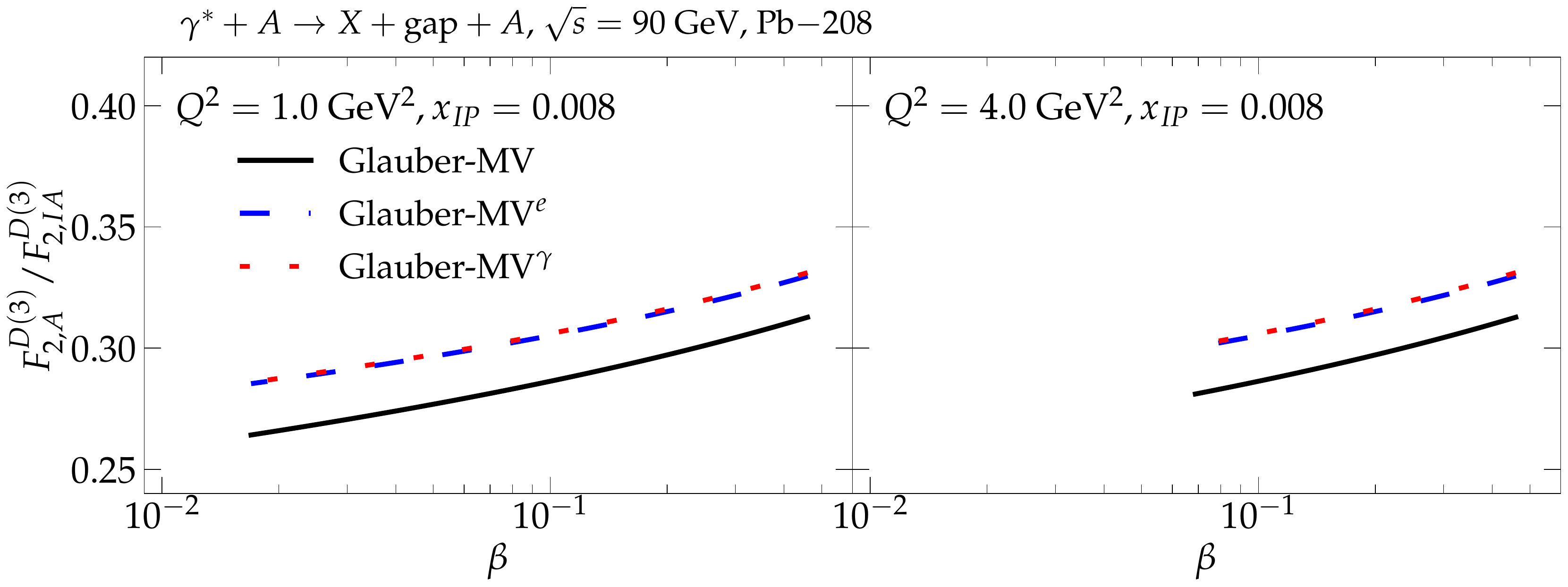}
    \caption{Nuclear modification ratio $F^{D(3)}_{2,A}/F^{D(3)}_{2,IA}$ as a function of $\beta$, when $\xpom$ is kept fixed and at two different values of the virtuality $Q^2$. Only numerics with $\beta<0.5$ are shown.}
    \label{fig:nucl_f2}
\end{figure}

The $\beta$ spectra of the diffractive structure function normalized by the full impulse approximation (IA) result, $F^{D(3)}_{2,A}/F^{D(3)}_{2,IA}$, where the latter can be obtained by setting $b_{\rm cut} = 0$ in~\cref{eq:large_b_cut}, are shown in Fig.~\ref{fig:nucl_f2}, when $\xpom$ is kept fixed. The most prominent prediction is a strong nuclear suppression at the chosen kinematics which is accessible at the future EIC. This can be explained by noting that, from Ref.~\cite{Hatta:2006hs} and Eqs.~(\ref{eq:large_b_cut}), at small enough $\xbj$, one can approximately estimate the cross-section ratio keeping only the leading behavior as:   
\begin{equation}
    \label{eq:nucl_mod_estimate}
    \frac{\sigma_{\rm D}^{\gamma^*A}}{\sigma_{\rm D, IA}^{\gamma^*A}} \sim \frac{\int \dd^2\bt \left[(Q_{s,A}^2(b)/Q^2\right]}{\sigma_0^2A^{4/3}\left[Q_{s,p}^2/Q^2\right]} \sim \sigma_0^{-1}A^{-1/3},
\end{equation}
where $Q_{s,A}$ and $Q_{s,p}$ are the nuclear and proton saturation scales, respectively. (A more careful estimation %taking into account the anomalous dimension from the BK evolution 
is available in Ref.~\cite{Lappi:2023frf} which  shows even a stronger suppression.) Fig.~\ref{fig:nucl_f2} also shows a weak decrease toward small $\beta$, which cannot be justified by \cref{eq:nucl_mod_estimate} given that $Q_{s,A}$ and $Q_{s,p}$ have a similar $\xbj$-dependence (recall that $\xbj=\xpom\beta$). Note however that such estimation is more reliable at asymptotically small $x$ and includes only the leading terms. The weak $\beta$ dependence is due to the presence of sub-asymptotic and/or subleading effects, which is more important at larger $\xbj$.       

\section{Conclusions}

We have reported a numerical study of the coherently diffractive dissociation for both proton and nuclear targets in the moderately high-mass regime within the kinematics of HERA and of the future EIC, respectively. The calculation employs the perturbative Kovchegov-Levin evolution equation, with the non-perturbative input being constrained from the HERA structure function data. The proton shape, which determines the overall normalization of the diffraction on the proton target, is optimized using the HERA combined diffractive data in the low-mass region.  
The KL evolution is shown to provide a reasonable description for the data in the desired region when a proton shape profile that is  steeper than the usual Gaussian profile is used. In the scattering off the nucleus, the results predict a strong nuclear suppression already for the kinematics accessible at the future EIC. 

The current results show the relevance of the KL evolution dynamics in the regime of interest, which is the transition region between low-mass and high-mass contributions. The question of a unified description of diffraction that smoothly connects these two extremes thus arises, and would be worthy of future investigation. 

\begin{acknowledgements}
This work was supported by the Academy of Finland, the Centre of Excellence in Quark Matter (project 346324)   and projects 338263 and 346567 (H.M), and 321840 (T.L). This work was also supported under the European Union’s Horizon 2020 research and innovation programme by the European Research Council (ERC, grant agreement No. ERC-2018-ADG-835105 YoctoLHC) and by the STRONG-2020 project (grant agreement No. 824093). The content of this article does not reflect the official opinion of the European Union and responsibility for the information and views expressed therein lies entirely with the authors. 
\end{acknowledgements}

\appendix
%\nocite*
\bibliographystyle{JHEP-2modlong.bst}
\bibliography{refs}
 
\end{document}